\documentclass[conference]{IEEEtran}
\IEEEoverridecommandlockouts
\usepackage{cite}
\usepackage{amsmath,amssymb,amsfonts}
\usepackage{algorithmic}
\usepackage{graphicx}
\usepackage{textcomp}
\usepackage{xcolor}
\usepackage{booktabs}
\usepackage{graphicx}
\usepackage{hyperref}

\def\BibTeX{{\rm B\kern-.05em{\sc i\kern-.025em b}\kern-.08em
    T\kern-.1667em\lower.7ex\hbox{E}\kern-.125emX}}
\makeatletter
\def\footnoterule{\relax
  \kern-3pt % Space above the rule
  \hbox to \columnwidth{\hfill\vrule width 3.5in height 0.4pt\hfill}
  \kern2.6pt} % Space below the rule
\makeatother
\begin{document}

\title{PathoWAve: A Deep Learning-based Weight Averaging Method for Improving Domain Generalization in Histopathology Images 

% \thanks{\textcolor{black}{This work was supported by the Natural Sciences and Engineering Research
% Council (NSERC) of Canada and in part by the Regroupement Stratégique en
% Microsystèmes du Québec (ReSMiQ).}}
}
\author{\IEEEauthorblockN{Parastoo Sotoudeh Sharifi, M. Omair Ahmad, \textit{Fellow, IEEE}, and M.N.S. Swamy, \textit{Fellow, IEEE}}
\IEEEauthorblockA{
\textit{Department of Electrical and Computer Engineering, Concordia University, Montreal, QC H3G 1M8, Canada}\\
\textit{Email: \{p\_sotoud, omair, swamy\}@ece.concordia.ca}}

}

\maketitle

\begin{abstract}
Recent advancements in deep learning (DL) have significantly advanced medical image analysis. In the field of medical image processing, particularly in histopathology image analysis, the variation in staining protocols and differences in scanners present significant domain shift challenges, undermine the generalization capabilities of models to the data from unseen domains, prompting the need for effective domain generalization (DG) strategies to improve the consistency and reliability of automated cancer detection tools in diagnostic decision-making. In this paper, we introduce Pathology Weight Averaging (PathoWAve), a multi-source DG strategy for addressing domain shift phenomenon of DL models in histopathology image analysis. Integrating specific weight averaging technique with parallel training trajectories and a strategically combination of regular augmentations with histopathology-specific data augmentation methods, PathoWAve enables a comprehensive exploration and precise convergence within the loss landscape. This method significantly enhanced generalization capabilities of DL models across new, unseen histopathology domains. To the best of our knowledge, PathoWAve is the first proposed weight averaging method for DG in histopathology image analysis. Our quantitative results on Camelyon17 WILDS dataset demonstrate PathoWAve's superiority over previous proposed methods to tackle the domain shift phenomenon in histopathology image processing. Our code is available at \href{https://github.com/ParastooSotoudeh/PathoWAve}{https://github.com/ParastooSotoudeh/PathoWAve}
\end{abstract}


\section{Introduction}

Histopathology is key in diagnosing and prognosticating diseases, especially in oncology. Whole slide images (WSIs) from tissue sections offer critical insights into tissue morphology, cellular structures, and disease progression. However, WSI analysis faces challenges due to variability in staining protocols, scanners, tissue preparation, and imaging systems across medical centers, causing significant data distribution changes. This variability leads to domain shift, where DL models trained on data from one domain falter in generalizing to unseen domains, compromising diagnoses in clinical applications~\cite{zhou2022domain}. DG techniques that enhance model invariance to data distribution changes, promising consistent performance across medical settings, are crucial for tackling the domain shift challenge. We categorized some of the current related proposed methods into the following categories: 

DG methods: CORAL~\cite{sun2016deep} aligns domain feature distributions via covariance matching, foundational to domain adaptation advances. IRM~\cite{arjovsky2019invariant} introduces a framework for learning domain invariances, enhancing generalization to unseen environments. Group DRO~\cite{sagawa2019distributionally} focuses on worst-case domain performance, improving model resilience. FISH~\cite{shi2021gradient} targets feature-level domain discrepancies for domain-invariant representation learning. PLDG~\cite{yan2024prompt} employs pseudo labeling for cross-domain data variability, enhancing model adaptability. TFS-ViT~\cite{noori2024tfs} utilizes token-level feature stylization in Vision Transformers for robustness, marking progress in domain generalization (DG) techniques. These methods highlight the progression towards bridging the source-target domain gap, fostering more refined approaches.

Data Augmentation methods: Data augmentation plays a pivotal role in DG. Data augmentation strategies~\cite{cubuk2020randaugment, zhou2020learning} provide simple yet effective strategies for introducing variability into the training process.\\
Test-Time Methods: Test-Time Training methods, like Test-time image-to-image translation~\cite{scalbert2022test}, offer promising way for dynamically adjusting models in response to new domain characteristics encountered at inference.

In this paper, we propose PathoWAve, a domain generalization technique, tailored specifically for histopathology image analysis inspired by recent advances in weight averaging methods like SWA~\cite{izmailov2018averaging}, SWAD~\cite{cha2021swad}, Lookahead~\cite{zhang2019lookahead}, and Lookaround~\cite{zhang2024lookaround} and the understanding of loss landscapes in neural networks, particularly the concept of Linear Mode Connectivity (LMC)~\cite{entezari2021role}. 
Our significant contributions through deploying PathoWAve on a histopathology dataset include:
\begin{itemize}
\item Introduction of PathoWAve for addressing domain shift in Histopathology: PathoWAve, a multi-source domain generalization method, tackles domain shift challenges in histopathology image analysis caused by varied staining techniques and imaging conditions. It enhances generalization by training identical models on diversely augmented images in parallel, integrating an advanced weight averaging strategy within the training cycle to ensure model diversity and locality.
\item Strategic Combination of regular and histopathology-specific augmentation methods: PathoWAve merges regular and histopathology-specific augmentation techniques, notably employing HEDJitter, a unique method for histopathology images~\cite{tellez2018whole}, demonstrating improved results and enhanced generalization through this combination.
\item State of the Art in weight averaging for DG in histopathology images: PathoWAve pioneers the use of weight averaging and combining histopathology-specific augmentation with regular augmentations to combat domain shift in histopathology images, marking a first in DG in histopathology analysis. Tested on the Camlyon17-WILDS dataset, our method outperforms existing DG techniques, proving its efficacy in mitigating domain shift and boosting model robustness to variations in unseen data.
\end{itemize}

\begin{figure}[!t]
\centering
\includegraphics[width=0.80\linewidth]{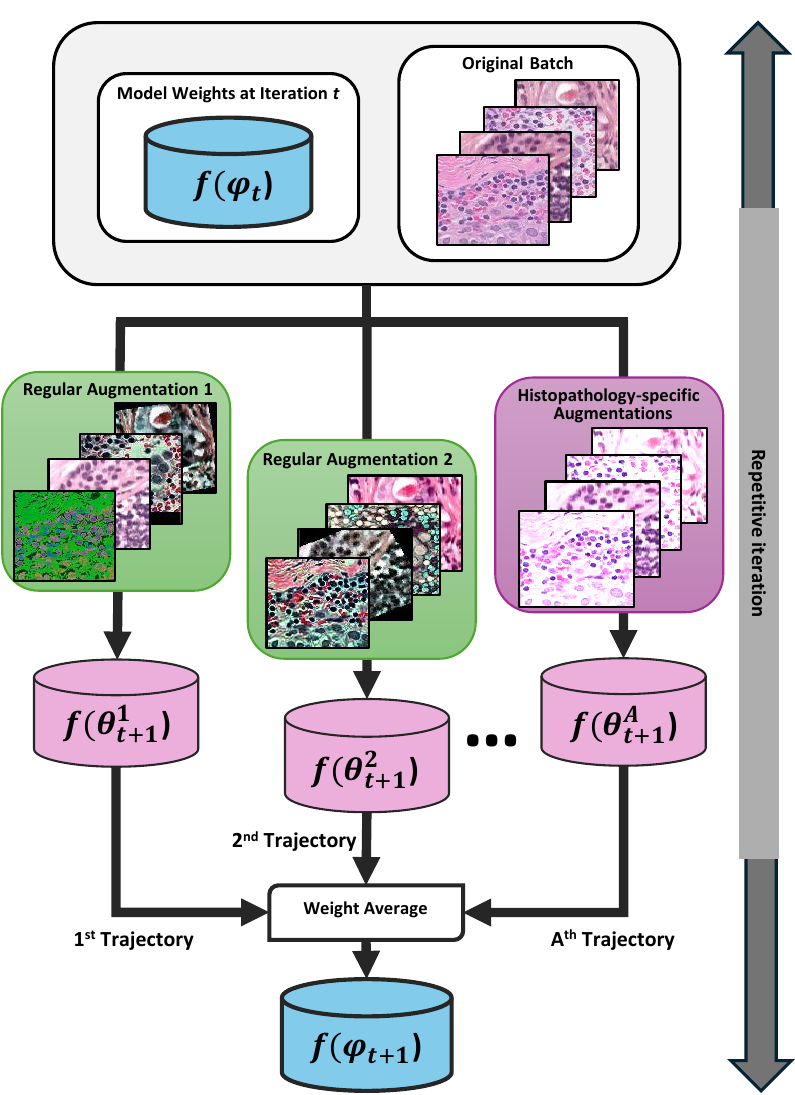}
\caption{\scriptsize Overview of the PathoWAve training process. This framework illustrates the cyclical weight averaging across multiple training trajectories, each applying a unique set of augmentations—regular and histopathology-specific—to an original batch of images.}

\label{fig:histo}
\end{figure}

\section{Method}
In the specific context of medical image processing, particularly histopathology image analysis, significant challenges arise from the variability in staining techniques, scanners, imaging conditions, and tissue processing methods. These factors contribute to domain shift challenges in real-world scenarios, where accurate and timely disease detection is critical for patient care. Consequently, there is a pressing need for robust domain generalization strategies that can effectively address these issues, ensuring models remain invariant to data distribution shifts and maintain their reliability in diagnostic decision-making. In the subsequent sections, we first delve into the intricacies of domain generalization, laying the groundwork for understanding its significance. Subsequently, we introduce and elaborate on our proposed method, PathoWAve, specifically designed to address domain generalization challenges in histopathology image analysis.

\subsection{Domain Generalization Objective}
Domain generalization (DG) tackles the challenge of developing models that, when trained on multiple source domains, exhibit robust performance on previously unseen target domains. Each domain is characterized by its unique joint distribution over input space $X$ and target space $Y$, denoted as $P_{X,Y}$. The aim is to leverage the diversity among source domains to predict accurately in target domains that the model has not encountered during training.

Formally, the DG framework seeks to minimize the expected loss over unseen target domains, $D_{target}$, using knowledge derived from a set of source domains. This objective can be expressed mathematically as:
\begin{equation}
\min_{f} \mathbb{E}_{(x,y) \in D_{target}}[L(f(x), y)],
\end{equation}
where \(f: X \rightarrow Y\) is the predictive function designed to approximate the posterior distribution \(P_{Y|X}\), and \(L\) is the loss function measuring the discrepancy between predicted and actual labels.

Given that direct access to $D_{target}$ is not feasible during training, the strategy shifts towards minimizing the empirical risk over the source domains, represented as:
\begin{equation}
\min_{\phi} \frac{1}{N} \sum_{k=1}^{N} L(f_{\phi}(x_k), y_k),
\end{equation}
where $N$ is the cumulative number of samples across all source domains, $\phi$ represents the parameters of model $f$, and $f_{\phi}$ signifies the model parameterized by $\phi$.

This problem setup acknowledges the inherent variability across domains by assuming each source domain $D_{source}^{i}$ for $i=1, \ldots, M$ has a distinct joint distribution $P_{X,Y}^{i}$, reflecting the real-world scenario where training data might not encompass the full spectrum of variation present in unseen target data. The ultimate goal of DG is to construct a model that, despite the distributional differences encapsulated by $P_{X,Y}^{i} \neq P_{X,Y}^{j}$ for $i \neq j$, can generalize effectively across novel domains, thereby ensuring reliable predictions for $D_{target}$ without requiring explicit knowledge of its distribution.

\subsection{PathoWAve Framework}
Our PathoWAve framework introduces a novel cyclical training regime that leverages the strengths of specific weight averaging method during training process and integrates both standard and histopathology-specific data augmentations. This strategic combination aims to cultivate an expansive exploration of both locality and diversity within the model's learning process, ultimately enhancing generalization capabilities across unseen histopathology domains. Figure~\ref{fig:histo} illustrates the overall architecture of our proposed PathoWAve framework.

Inspired by recent advancements in weight averaging methods~\cite{izmailov2018averaging, cha2021swad, zhang2024lookaround, zhang2019lookahead}, we explore the potential of a multi-training trajectory approach. By having multiple training trajectories on an identical model architecture but diversified through carefully selected data augmentations, including histopathology-specific ones designed to simulate a broad spectrum of staining variations and imaging conditions, this approach seeks to achieve beneficial model diversity within the scope of histopathology images and thus enhancing generalization to unseen domains.

More specifically, our method employs the concurrent training of multiple identical neural networks, each initialized from a common point within the loss landscape. Each training path will be exposed to distinct batches ($B$) of data, with each batch $B$ sampled from the union of all source domain datasets $\bigcup_{i=1}^{M} D_{source}^{i}$. These batches are processed through a tailored suite of data augmentation strategies ($AUG^n$) to introduce a broad spectrum of variations reflective of real-world histopathological conditions.

The training process for each network \(i\) is mathematically captured by the update formula:

\begin{equation}
\theta_{t+1}^{n} = \phi_{t} + \eta \cdot \nabla_{\phi} \mathcal{L}(\phi_{t}, AUG^n(B)),
\end{equation}

where $\theta_{t+1}^{n}$ represents the model parameters for the \(n^{th}\) network updated after training iteration $t$, $\eta$ denotes the learning rate, and $\nabla_{\phi} \mathcal{L}(\phi_{t}, AUG^n(B))$ signifies the gradient of the loss function $\mathcal{L}$ with respect to the averaged model parameters $\phi_{t}$, evaluated on the augmented batch $AUG^n(B)$. This approach ensures that each network \(n\) is exposed to a variety of data representations through its specific augmentation strategy $AUG^n$, enhancing the overall diversity of the model's learning experience and its ability to generalize across unseen domains.

Following this phase of individual training, we integrate the weights of models of each training path through a weight averaging strategy, applied directly within the training cycle, aiming to reach flatter minima within the loss landscapes which results in better generalization of the model. This process is formalized as follows:

\begin{equation}
\phi_{t+1} = \frac{1}{A} \sum_{n=1}^{A} \theta_{t+1}^{n},
\end{equation}

where $\phi_{t+1}$ represents the unified set of weights obtained by averaging the parameters $\theta_{t+1}^{n}$ of each model $n$ out of the total $A$ models at iteration $t+1$. This integration forms a cohesive weight set, serving as the starting point for all models in subsequent iterations and substantially enhancing their generalization capabilities. In this way, in addition to improving the diversity by having a multi-training trajectory and using specific augmentation methods, PathoWAve facilitates the convergence of the models towards lower-loss regions during the whole training cycle, promoting model robustness and locality. Through the PathoWAve method, we achieve a sophisticated implementation of the DG objective, effectively minimizing loss across unseen domains via a structured, iterative refinement process.

Choosing suitable augmentations plays a crucial role in the efficacy of the PathoWAve framework. Our approach uniquely combines general data augmentations—such as AutoAugment and RandAugment~\cite{cubuk2020randaugment}—with histopathology-specific augmentations like HEDJitter~\cite{tellez2018whole} to address the broad spectrum of variability encountered in histopathological images. General augmentations introduce a wide range of variations in the dataset, fostering the model’s adaptability and robustness against common variations in image data. In contrast, the HEDJitter augmentation technique is meticulously designed for histopathology images, utilizing a predefined Optical Density (OD) matrix to transition images from RGB to a domain that emphasizes pathology stains—hematoxylin, eosin, and Diaminobenzidine (DAB). This technique's capacity to independently adjust stain intensity levels simulates the diverse staining protocols found across laboratories, ensuring the preservation of crucial image features such as cell structures and tissue architecture.

The integration of both regular and histopathology-specific augmentations into the PathoWAve training regime is a strategic decision aimed at enhancing the model's exposure to a wide array of data variations, thereby ensuring a more comprehensive learning experience. Regular augmentations prepare the model for a broad base of image variations, enhancing its adaptability and resilience to general shifts in input data distributions. Meanwhile, HEDJitter and similar histopathology-specific techniques target the nuanced challenges specific to histopathological imagery, such as staining variability, which are critical for achieving high diagnostic accuracy in unseen domains. This dual-strategy augmentation approach not only widens the model's exposure but also fine-tunes its sensitivity to the unique challenges of histopathology image analysis. Consequently, PathoWAve exhibits outstanding generalization capabilities, setting a new benchmark in domain generalization for histopathology image analysis by leveraging this comprehensive augmentation strategy.

\section{Experiments}
\textbf{Dataset:}
In our experiments, we used the Camelyon17 WILDS dataset~\cite{koh2021wilds}, featuring patches from Whole Slide Images of lymph node sections across five medical centers with diverse staining protocols and scanners, to test DL models on metastatic breast cancer detection. This dataset is partitioned by medical center origin, for developing generalized models to unseen data for cancerous tissues detection. For training and identification validation (id val), data come from three hospitals (30 WSIs and 302,436 patches for training, plus 33,560 patches for id val), while validation (val) and testing datasets are sourced from unique, previously unseen hospitals—val with 10 WSIs and 34,904 patches from one hospital, and testing with 10 WSIs and 85,054 patches from another hospital, ensuring models are assessed on their adaptability to new medical center data.\\
\textbf{Implementation Details:} 
 We utilized the ResNet50 as our network, and we used an NVIDIA V100 32 GB GPUs for all of our experiments. The learning rate and batch size are set to $2e-5$ and $128$, respectively.  To enhance model's robustness to staining variations, we incorporated several augmentation methods, including HedJitter augmentation~\cite{tellez2018whole} with a jitter\_strength of 0.05.

\begin{table}[!t]
    \centering
    \caption{\scriptsize Comparative performance of various domain generalization methods on validation and test set of Camelyon17 WILDS dataset. $^\dag$ and $^\ddag$ results are from \cite{scalbert2022test} and \cite{yan2024prompt}, with remaining methods from their corresponding original studies.}
    \label{tab:dg} % Label for referencing
    \resizebox{0.85\linewidth}{!}{%
    \begin{tabular}{@{}lccc@{}}
    \toprule
        Method & Backbone & Validation \% & Test \% \\ 
    \midrule
        CORAL$^\dag$ (2016)~\cite{sun2016deep} & ResNet50 & 86.2 & 59.5 \\ 
        IRM$^\dag$ (2019)~\cite{arjovsky2019invariant} & ResNet50 & 86.2 & 64.2 \\ 
        Group DRO$^\dag$ (2019)~\cite{sagawa2019distributionally} & ResNet50 & 85.5 & 68.4 \\ 
        DomainMix (2020)~\cite{xu2020adversarial} & ResNet50 & --- & 69.7 \\ 
        MMLD$^\ddag$ (2020)~\cite{matsuura2020domain}\ & ResNet50 & --- & 70.2 \\
        ERM (2021)~\cite{koh2021wilds} & ResNet50 & --- & 70.3 \\ 
        FISH$^\dag$ (2021)~\cite{shi2021gradient} & ResNet50 & 83.9 & 74.7 \\ 
        V\_REx (2021)~\cite{krueger2021out} & ResNet50 & --- & 71.5 \\ 
        IB-IRM  (2021)~\cite{ahuja2021invariance} & ResNet50 & --- & 68.9 \\ 
        LISA (2022)~\cite{yao2022improving} & ResNet50 & --- & 77.1 \\ 
        FuseStyle (2023)~\cite{khamankar2023histopathological} & ResNet50 & --- & 90.5 \\ 
        \midrule
        CORAL$^\ddag$ (2016)~\cite{sun2016deep} & ViT-Base & --- & 71.8 \\
        DANN$^\ddag$ (2016)~\cite{ganin2016domain} & ViT-Base & --- & 83.5 \\
        IRM$^\ddag$ (2019)~\cite{arjovsky2019invariant} & ViT-Base & --- & 75.0 \\
        ERM$^\ddag$ (2021)~\cite{koh2021wilds}& ViT-Base & --- & 73.1 \\
        SelfReg$^\ddag$ (2021)~\cite{kim2021selfreg} & ViT-Base & --- & 70.4 \\
        PLDG$^\ddag$ (2024)~\cite{yan2024prompt} & ViT-Base & --- & 84.3 \\
        EPVT$^\ddag$ (2024)~\cite{yan2023epvt} & ViT-Base & --- & 86.4 \\        
        \midrule
        % \multicolumn{4}{@{}l@{}}{\textbf{Train Time Augmentation Methods}} \\ 
        % \textcolor{red}{RandAugment} (2019)~\cite{cubuk2020randaugment} & ResNet50 & 90.6 & 82.0 \\ 
        % StarGanV2 data \textcolor{red}{aug} & ResNet50 & 89.6 & 76.4 \\ 
        \textbf{PathoWAve (ours)} & \textbf{ResNet50} & \textbf{93.07} & \textbf{94.36} \\ 
    \bottomrule
    \end{tabular}%
    }
\end{table}

\begin{table}[!t]

    \centering
    \caption{\scriptsize Performance comparison of PathoWAve against non-DG methods on Camelyon17 WILDS.}
    \label{tab:non_dg} % Label for referencing
    \resizebox{1.0\linewidth}{!}{%
    \begin{tabular}{@{}llcc@{}}
    \toprule
        Method & Description & Validation \% & Test \% \\ 
    \midrule        
        % RandAugment (2019)~\cite{cubuk2020randaugment} & ResNet50 & 90.6 & 82.0 \\ 
        STRAP (2021)~\cite{yamashita2021learning} & Uses source art images & --- & 93.7 \\ 
        StarGanV2 (2022)~\cite{scalbert2022test} & Train-time data augmentation & 89.6 & 76.4 \\ 
        TestTimeI2I (2022)~\cite{scalbert2022test} & Test-time adaptation & 92.8 & 94.0 \\ 
        \midrule
        \textbf{PathoWAve (ours)} & \textbf{Domain Generalization} & \textbf{93.07} & \textbf{94.36} \\ 
    \bottomrule
    \end{tabular}%
    }
\end{table}

\section{Results}
\textbf{Comparison with State of the Art:} 
Our comprehensive evaluation on the Camelyon17 WILDS dataset, presented in Tables~\ref{tab:dg} and \ref{tab:non_dg}, illustrates PathoWAve's exceptional capability to generalize across domain shifts within histopathology images. The comparison includes robust domain generalization (DG) methods, underlining PathoWAve's state-of-the-art performance. Remarkably, PathoWAve, leveraging a straightforward ResNet architecture, excels beyond more complex architectures, including those based on the vision transformer (ViT). This underscores the efficiency of our proposed method. Moreover, a comparison with non-DG methods, including advanced training-time augmentation and test-time adaptation techniques, further highlights PathoWAve's effectiveness. Specifically, PathoWAve outperforms methods like STRAP, which leverages non-histopathological data, and others employing dynamic adaptations during test time, as shown in our comparisons. Crucially, PathoWAve attains this high level of generalization and accuracy without leveraging direct test data insights, emphasizing the robustness of our proposed approach.

\begin{table}[!t]
    \centering
    \caption{\scriptsize Ablation analysis of PathoWAve method showing the impact of different combinations of augmentations and the number of independent training trajectories on the Camelyon17 WILDS dataset.}
    \label{tab:abl} % Label for referencing
    \resizebox{1.0\linewidth}{!}{%
    \begin{tabular}{@{}llc@{}}
    \toprule
        Method & \#Independent Trajectories (Augmentations) & Test \% \\ 
    \midrule        
        ERM & 1 (baseline with no weight averaging) & 70.3 \\ 
        \midrule
        PathoWAve & 2 (AutoAugment, RandomAugment) & 92.53 \\ 
        PathoWAve & 2 (RandomAugment, HEDJitter) & 92.98 \\
        PathoWAve & 2 (AutoAugment, HEDJitter) & 94.20 \\
        \midrule
        PathoWAve & 3 (AutoAugment, RandomAugment, AutoRandomRotation) & 89.80 \\
        PathoWAve & 3 (AutoAugment, RandomAugment, RandomGaussBlur) & 88.91 \\
        PathoWAve & 3 (AutoAugment, RandomAugment, RandomAffine) & 91.53 \\
        PathoWAve & 3 (AutoAugment, RandomAugment, HEDJitter) & \textbf{94.36} \\
    \bottomrule
    \end{tabular}%
    }
\end{table}

\textbf{Ablation Analysis:} 
Our detailed ablation study, as summarized in Table~\ref{tab:abl}, evaluates the impact of various augmentation strategies and the number of independent training trajectories on the PathoWAve method's effectiveness within the domain of histopathology image analysis on the Camelyon17 WILDS dataset. Initially establishing a baseline with the ERM method, which utilizes a single training trajectory without weight averaging, yielded a test accuracy of $70.03\%$. The introduction of PathoWAve with dual augmentation strategies significantly enhances model performance, highlighting the method's responsiveness to diverse training signals.

Notably, combinations involving two augmentations, particularly AutoAugment with HEDJitter, demonstrated remarkable improvements, achieving a test accuracy of $94.20\%$. This underscores the critical role of HEDJitter, a histopathology-specific augmentation, in bolstering the model's generalization capability across unseen domains.

Further exploration with three augmentations revealed varying degrees of success. While adding AutoRandomRotation, RandomGaussBlur, or RandomAffine to the AutoAugment and RandomAugment mix led to lower test accuracies compared to dual-augmentation setups, the incorporation of HEDJitter alongside AutoAugment and RandomAugment within a three-trajectory framework achieved the highest performance at $94.36\%$. This pinnacle result not only signifies the optimal augmentation combination but also establishes PathoWAve as the state-of-the-art in domain generalization for histopathology images.

It is worth mentioning that our proposed method's training time is A times that of traditional one-trajectory methods, as we perform A augmentations in parallel per iteration before weight averaging. Importantly, this overhead is only during training; the testing time remains the same as other methods since we use the averaged weights for evaluation.

\section{Conclusion}
% Our study presents PathoWAve, a weight averaging methodology for DG in histopathology imaging, achieving significant domain shift mitigation. Utilizing a multi-trajectory training strategy and a tailored mix of histopathology-specific augmentation with other augmentation techniques, PathoWAve outperforms existing models in robustness and accuracy on the Camelyon17 WILDS dataset, demonstrating its potential to improve automated cancer detection. Future research will explore additional histopathology-specific augmentations to further enhance model generalization, aiming to develop more robust and reliable diagnostic tools for cancer detection. Additionally, We aim to collaborate with other real-world benchmarks to access a broader array of histopathological images. This diversity will help in fine-tuning the model to handle practical variations in medical imaging conditions more effectively.

Our study presents PathoWAve, a weight averaging methodology for DG in histopathology imaging, achieving significant domain shift mitigation. Utilizing a multi-trajectory training strategy and a tailored mix of histopathology-specific augmentation with other augmentation techniques, PathoWAve outperforms existing models in robustness and accuracy on the Camelyon17 WILDS dataset, demonstrating its potential to improve automated cancer detection. Future research will explore the performance of our method on other real-world benchmarks in addition to Camelyon17, as well as additional histopathology-specific augmentations to further enhance model generalization. This aims to develop more robust and reliable diagnostic tools for cancer detection.
% \section*{References}
\bibliographystyle{IEEEtran}
\bibliography{egbib}

\end{document}